\documentclass[12pt]{article}

 \usepackage{amsfonts}
 \usepackage{amssymb,amsmath}
 \usepackage{graphicx} \usepackage{epstopdf}
 \newcommand{\crlb}[1]{\label{#1}\\[2pt]}
 
 \newcommand{\crld}[1]{\label{#1}}
 \newcommand{\eela}[1]{\quad\hbox{\scriptsize{#1}}\label{#1}\end{eqnarray}}
 \newcommand{\eelb}[1]{\label{#1}\end{eqnarray}}
 
 \newcommand{\newsecb}[2]{\section{#1}\label{#2}\setcounter{equation}{0}}

 \newcommand{\nolabels} {\def\eel{\eelb}\def\eeql{\eeqlb}  \def\crl{\crlb} 
 \def\newsecl{\newsecb}\def\bibiteml{\bibitem} \def\citel{\cite}\def\labell{\crld}}
\newcommand{\eeqla}[1]{\quad\hbox{\scriptsize{#1}}\label{#1}\end{aligned}\end{equation}}
\newcommand{\eeqlb}[1]{\label{#1}\end{aligned}\end{equation}}

   \def\Intechversion{\nolabels\setlength{\oddsidemargin}{.2in}\setlength
    {\textwidth}{5.87in} \setlength{\topmargin}{-.5in}\setlength\textheight{8in}}
\def\beq{\begin{equation}\begin{aligned}}		\def\eeq{\end{aligned}\end{equation}}
\def\be{\begin{eqnarray}}  					\def\ee{\end{eqnarray}}		
   \def\bi#1{\begin{itemize}\item[#1]} 	      	   \def\ei{\end{itemize}} 
   \def\eqn#1{(\ref{#1})}
   	 \def\fn{\footnote}	 

		 \def\alf{\alpha}   \def\bet{\beta}   \def\del{\delta}        \def\lab{\lambda}

	    		        		     \def\vv{\varphi}     
 	 		\def\s{\sigma}     	      	 
	     		          		              	  
    		  		\def\dd{{\rm d}} 		\def\HH{{\mathcal H}}

\def\bra{\langle} 		\def\ket{\rangle}

\def\qu{\overset{\textstyle ?}{=}}

\def\fract#1#2{{\textstyle\frac{#1}{#2}}}	 	 	
\def\ffract#1#2{\raise .2 em\hbox{$\scriptstyle#1$}\kern-.25em/\kern-.35em\lower .15 em \hbox{$\scriptstyle\,#2$}}
			\def\halff{\ffract12}		
 
\def\ex#1{e^{\textstyle#1}} 		\def\qqquad{\qquad\qquad}	
\def\bpmatrix{\begin{pmatrix}} 			\def\epmatrix{\end{pmatrix}}
\def\bmatrix{\begin{matrix}} 			\def\ematrix{\end{matrix}} 
\def\bcenter{\begin{center}}			\def\ecenter{\end{center}}

\def\lowerheightfig#1#2#3{\(\raise-#1\hbox{\includegraphics[height=#2]{#3}}\)}
\def\lowerwidthfig#1#2#3{\(\raise-#1\hbox{\includegraphics[width=#2]{#3}}\)}
\def\widthfig#1#2{\includegraphics[width=#1]{#2}}
\def\th{\({}^{\mathrm{th}}\)}		 
\def\intt{{\mathrm{int}}}

\def\ol{\overline}  
\def\hok{\(\hbox{ }\)}  \def\ont{{\mathrm{ont}}}
 \def\twomat#1#2{\Big(\begin{matrix} #1 \\ #2 \end{matrix}\Big)} %\twomat{a&b&c}{&d&} etc.
\def\weglaten#1{}	
  \Intechversion	% see settings chosen above. Change as desired.

\begin{document}
							\begin{titlepage}
 \title{ \Huge\bf  Ontology in quantum mechanics
\author{Gerard 't~Hooft}
\date{\normalsize
Faculty of Science,
Institute for Theoretical Physics,\\
Utrecht University, \\
Princetonplein 5,
3584 CC Utrecht, \\
\underline{The Netherlands} \\[10pt]
e-mail:  g.thooft@uu.nl \\ internet: 
http://www.staff.science.uu.nl/\~{}hooft101/ }} \maketitle
 
 {\noindent \large \textbf{Abstract}}\\[10pt]
   \indent It is suspected that the quantum evolution equations describing the micro-world as we know it are of a special kind that allows transformations to a special set of basis states in Hilbert space, such that, in this basis, the evolution is given by elements of the permutation group. This would restore an ontological interpretation. It is shown how, at low energies per particle degree of freedom, almost any quantum system allows for such a transformation. This contradicts Bell's theorem, and we emphasise why some  of the assumptions made by Bell to prove his theorem cannot hold for the models studied here. We speculate how an approach of this kind may become helpful in isolating the most likely version of the Standard Model, combined with General Relativity.  A link is suggested with black hole physics. \\[10pt]
    
\noindent\textbf{Keywords:}\  foundations quantum mechanics,
fast variables,
cellular automaton,
classical/quantum evolution laws, Stern-Gerlach experiment,
Bell's theorem, free will,
Standard Model,
anti-vacuum state.\\
 
	\end{titlepage}
 \newsecl{Introduction}{intro}	%\newsecl shows label, here: {intro}, when testprintversion is used
Since its inception, during the first three decades of the 20\th century, quantum mechanics was subject of intense discussions concerning its interpretation. Since experiments were plentiful, and accurate calculations could be performed to compare the experimental results with the theoretical calculations, scientists quickly agreed on how detailed quantum mechanical models could be arrived at, and how the calculations had to be done.

The question what the intermediate results of a calculation actually tell us about the physical processes that are going on, remained much more mysterious. Opinions diverged considerably, up to today, one hundred years later.

The historical events that led to this situation are well-known, and have been recounted in excellent reports\,\cite{Pais1986}; there is no need to repeat these here extensively. It was realised that all oscillatory motion apparently comes in energy packets, which seem to behave as particles, and that the converse should also be true: all particles with definite energies must be associated to waves. The original descriptions were somewhat vague, but the year 1926 provided a new landmark: Erwin Schr\"odinger's equation\,\cite{Schroedinger1926}. Originally, the equation  was intended to describe just one particle at the time, but soon enough it was found how to extend it to encompass many particles that may be interacting.

Indeed, in his original paper, Schr\"odinger went quite far in discussing Hamilton's principle, boundary conditions, the hydrogen atom and the electromagnetic transitions from one energy level to an other. One extremely useful observation was made by Max Born\,\cite{Born1926}: the absolute square of a wave function, at some spot in position space, must simply stand for the \emph{probability} to find the particle there. This made a lot of sense, and it was rightly adopted as a useful recipe for dealing with the equation.

But then, many more questions were asked, many of them very well posed, but the answers sound too ridiculous to be true, and, as I shall try to elucidate, they \emph{are} too ridiculous to be true.
I am not the only scientist who feels taken aback by the imaginative ideas that were launched, ranging from the role of `guiding pilot' adopted by the wave function\,\cite{Bohm1952} to steer particles in the right direction, to the idea that infinitely many `universes' exist, all forming parts of a more grandiose concept of `truth' called `multiverse' or `omniverse', an idea now known as the `many worlds interpretation'\,\cite{Everett1957,DeWitt1973}.

In contrast, an apparently quite reasonable conclusion was already reached in discussions among scientists in the 1920s, centred around Niels Bohr in Copenhagen, called the `Copenhagen Interpretation'. They spelled out the rules for formulating what the equations were, and how to elaborate them to make firm predictions. 
Indeed, we know very well how to use the equation. The properties of atoms, molecules, elementary particles and the forces between all of these can be derived with perplexing accuracy using it. The way the equation is used is nothing to complain about, but what exactly does it say? 

Paul Dirac for instance, advised not to ask questions that cannot be answered by any experiment; such questions cannot be important. We know precisely how to use Schr\"odinger's equation; all that scientists have to do is insert the masses and coupling parameters of all known particles into the equation, and calculate. What else can you ask for? Many of my colleagues decided to be strictly `agnostic' about the interpretation, which is as comfortable a position to take as what is was for 19\th century scientists to stay `agnostic' about the existence of atoms.

The Copenhagen verdict was: 
\bi{} ``There are many questions whose answers will not be in the range of any experiment to check; there will be no unanimous agreement on the interpretation of the equations, so stop asking."\ei

The present author accepts all conclusions the Copenhagen group had reached, except this last one. It will be important to ask for models that can elucidate the events that take place in an experiment. We do wish to know which sensible models can in principle explain the Schr\"odinger equation and which will not.

What happens to its wave function when you actually observe a particle? What does it mean if the Schr\"odinger equation suggests that interference takes place between different possible paths a particle can take?
Those questions I can now answer, but others are still way out of reach: the masses and coupling parameters of the elementary particles have been determined by experiment, but we do not have acceptable theories at all to explain or predict their values. If the history of science is something to be taken to mind, it may be that asking unconventional questions will lead to better insights.

The Schr\"odinger equation is simple and it works, but some of the explanations why it works seem to get the proportions of a Hieronymus Bosch painting. This does not sound right.  Almost a full century has passed since the equation was written down, and we still do not know what or whom to believe, while other scientists get irritated by all this display of impotence\,\cite{NGvK1988}. Why is it that we still do not agree?

I think  I know some of the answers, but almost \emph{everyone} disagrees with me. I have reached the conclusion that quantum mechanics indeed describes a completely deterministic world. Admittedly, I will leave some questions unanswered. The origin of the symmetries exhibited by the equations is not well understood. More advanced mathematics will have to be employed to answer such questions, as will be explained. Sharpening the scope of my claim, the point is that there is no mystery with quantum mechanics itself. Just leave questions concerning symmetries aside for the time being. In contrast with what others proclaimed, there is no logical conflict. This will be explained (section~\ref{SM.sec}).

What \emph{are} those masses and coupling strengths? Do particles exist that we have not yet been able to detect? Isn't it the scientist's job to make predictions about things we have not yet been able to unravel? These are questions that are  haunting us physicists. We have arrived at a splendid theory that accounts for almost anything that could be observed experimentally. It is called the \emph{Standard Model} of the subatomic particles. But this model also tells us that particles and forces may exist that we could not have detected today. Can we produce any theory that suggests what one might be able to find, in some distant future? And as of all those particles and forces that we do know about, is there a theory that \emph{explains} all their details? 

Today's theories give us little to proceed further from where we are now. The Standard Model explains a lot, but not everything. This is why it is so important to extend our abilities to do experiments as far as we can. Recently, audacious plans have been unfolded by the European particle physics laboratory CERN, for building a successor of its highly successful Large Hadron Collider (LHC). While the experimental groups working with the LHC have provided for strong evidence supporting the validity of the Standard Model up to the TeV domain, theoreticians find it more and more difficult to understand why this model can be all there is to describe what happens further beyond that scale. There must be more; our present theoretical reasoning leads to questions doubting the extent to which this model can be regarded as `natural'  if more of the same  particles at  higher  energies are allowed to exist, while the existence of totally new particles would be denied. 

Inspired by what historians of science are telling us about similar situations in the past history of our field, investigators are hoping for a `paradigm shift'. However, while it is easy to postulate that we `are doing something wrong', most suggestions for improvement are futile; suggesting that the Standard Model would be `wrong' is clearly not going to help us. The `Future Circular Collider' is a much better idea; it will be an accelerator with circumference of about 100 km, being able to reach a c.m.~collision energy of 100 TeV. The importance of such a device is that it will provide a crucial background forcing theoreticians to keep their feet on the ground: if you have a theory, it better agree with the newest experimental observations.

\newsecl{The generic realistic model}{realmodel.sec} 
The central core of our theory consists of a set of models whose logic is entirely classical and deterministic. Deterministic does not mean \emph{pre-}deterministic: there is no shortcut that would enable one to foresee any special feature of the future without performing extremely complex simulation calculations using the given evolution laws. There is no `conspiracy'. Also, we do not take our refuge into any form of statistics. The equations determine exactly what is happening. Of course we do not know today exactly what the equations are, but we do assume them to exist. 
\\[30pt]
\hok\qquad Figure 1. \\
\hok\qquad Generic evolution law\\ 
\hok\qquad  for  a realistic model. with\\
\hok\qquad different periodicities.\\
\hok\qquad In this example we see\\
\hok\qquad 5 cycles, with ranks \(2,\,3,\,6,\,8\) and \(11\).\\[-105pt]
 \({\vphantom{T}}\)\qqquad\qqquad \qqquad \qqquad\qquad
\lowerwidthfig{10 pt} {200 pt}{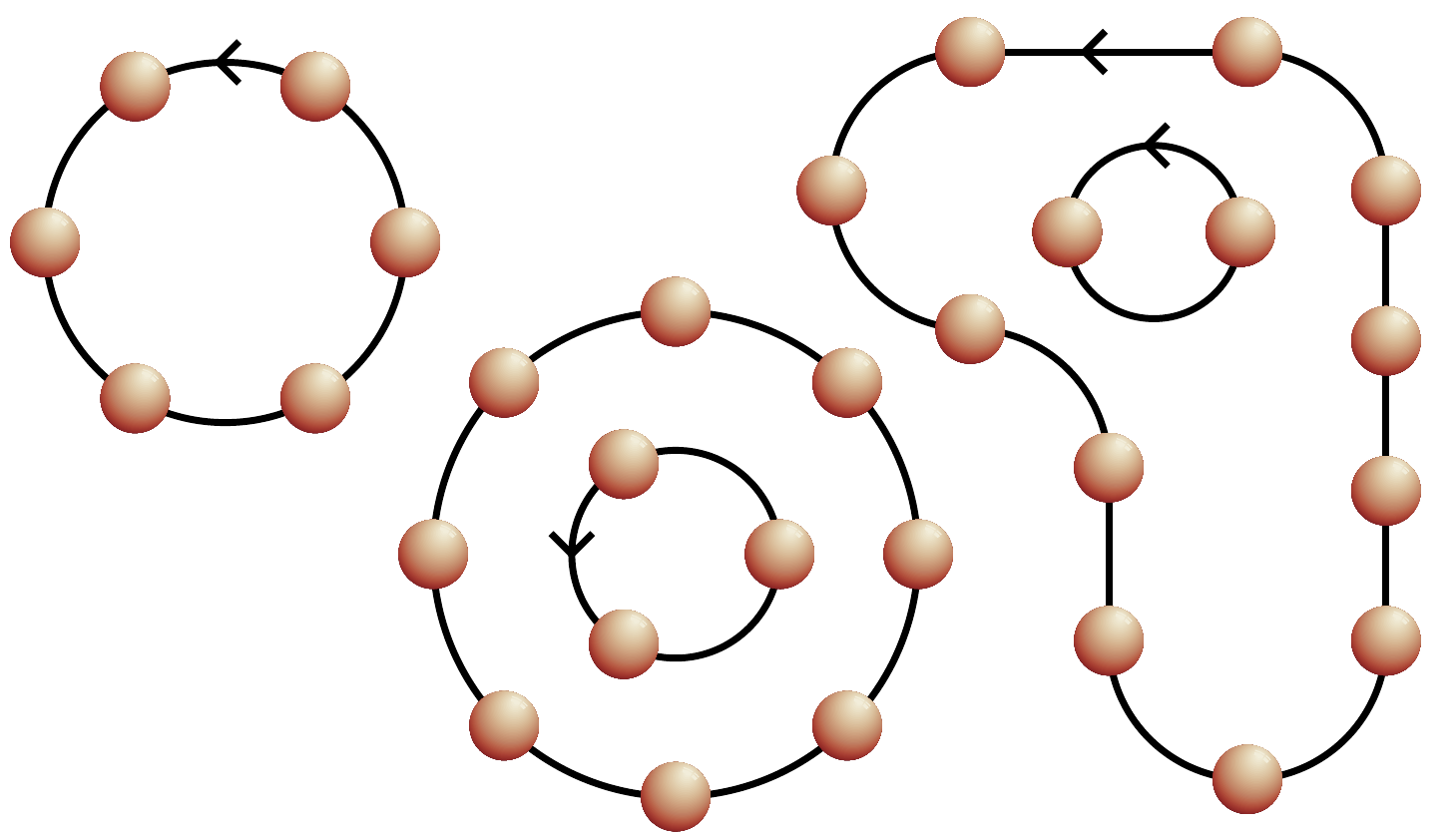}\\[10pt]

The equations will be \emph{more precise} even than Newton's equations for the motion of the planets. Newton's equations are given in terms of variables whose values are determined by real numbers. But, in practice, it is impossible to specify these numbers with infinite precision, and consequently, \emph{chaos} takes place: it is fundamentally impossible, for instance, to predict the location of the dwarf planet Pluto, one billion years from now, because such a calculation would require the knowledge of the locations and masses of all planets in more than 20 digits accuracy today\,\cite{SussWis1988}. That's a tiny fraction of a micron for Pluto's orbit. Following Pluto during the age of the universe would require accuracies beyond 2000 digits, much tinier margins than the Planck length. To describe Pluto's position exactly would require an infinite number of decimal places to be rigorously defined.

Our deterministic theory will be formulated in terms of integer numbers only, which can be defined exactly without the need of infinitely many decimal places. This kind of precision in defining theories may well be what is needed to understand quantum mechanics.

For simplicity, we imagine a universe with finite size and finite time.
As for their mathematical structure, all deterministic models are then very much alike. All finite-size discrete models must have finite Poincar\'e recursion times. There will be different closed cycles with different periods, see Figure 1. Counting these cycles, one finds that the rank of a cycle is physically a conserved quantity, almost by definition. For simplicity, we constrain ourselves to \emph{time reversible} evolution laws, although it is suspected that one might be allowed to relax this rule, but then the mathematics becomes more complex.

We now emphasise that the evolution law of such a deterministic system can be exactly described in terms of a legitimate, conventional Schr\"odinger equation. We say that quantum mechanics is a \emph{vector representation} of our model: every possible state the system can be in is regarded as a vector in the basis of Hilbert space. This set of vectors is ortho\-normal. The classical evolution law will send any of these vectors into an other one. Since these vectors are all ortho\-normal and since the evolution is time-reversible, one can easily prove that the evolution matrix is unitary. It contains only the numbers 1 and 0. There is only one 1 in each row and in each column; all the other entries are 0, from which unitarity follows.

By diagonalising this matrix, one finds all its eigenvectors and eigenvalues. Within one cycle, the eigenvalues of \(U(t)\) are
 \(e^{-2\pi i n t/T}\), where  \(t\) is time,  \(T\) is the period, and \(n\) is an integer. The formal expression for the eigenvectors is easily obtained: 
 \be {}^\ont\bra k|n\ket^E=\frac {1}{\sqrt N}\,\ex{-2\pi ink/N}\ , \eel {energyeigenstates.eq}
 where \(|k\ket^\ont\) are the ontological states, labelled by the integer \(k\), and \(|n\ket^E\) are the energy eigenvectors.
 We read off in the basis formed by the states \(|n\ket^E\) that the Hamiltonian takes the values
 \be H_{nm}=2\pi n\,\del_{nm}/T\,.\eel{eigenvaluesH.eq}
 
 At first sight, this does \emph{not} look like quantum mechanics; the series of eigenvalues  \eqn{eigenvaluesH.eq} seems to be too regular. In ref.~\cite{GtHCA} it was proposed to add arbitrary additive energy renormalization terms, depending on the cycle we are in, but the problem is then still that it is difficult to see how this can reproduce Hamiltonians that we are more familiar with. The energy eigenstates seem to consist of large sequences of spectral lines with uniform separations. A more powerful idea has been proposed recently\,\cite{GtH2020, GtH2021}. More use must be made of locality. We wish the Hamiltonian to be the sum of locally defined energy density operators, \be H=\sum_{\vec x}\HH(\vec x)\ . \ee
  Now this is really possible. The price to be paid is to add \emph{fast fluctuating, localised variables}, called `fast variables' for short. They replace the vague `hidden variables' that were introduced in many earlier proposals\,\cite{Neumann1932}.
 
 The fast variables, \(0\le\vv_i(\vec x)<2\pi\), are basically fields that rapidly repeat their values with periodicities 
 \(T_i(\vec x)\), which we choose all to be large and mostly different. To reproduce \emph{realistic} quantum mechanical models, we need these periods to be considerably shorter in time than the inverse of the highest energy collision processes that are relevant.
 
 To a good approximation, the fast variables will be non-interacting. This means that the energy levels will take the form \(E=2\pi\sum_{i,\vec x} n_i(\vec x)/T_i\), where the \(n_i\) are all integer, and it implies that there is one ground level, \(E_0=0\), while all excited states have energies \(E\ge 2\pi /T_i\). Clearly, our conditions on the fast variables were chosen such that their excited energy levels exceed all energy values that can be reached in our experiments.\fn{More precisely, we talk of energies that can be associated to single quantum particles at isolated points in space-time.}
 
\hok\qquad\widthfig{350pt}{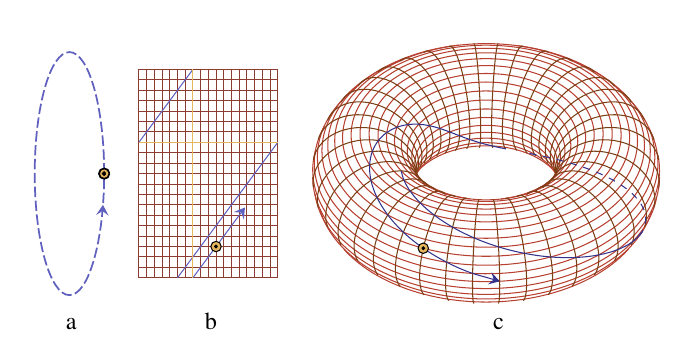}\\[10pt]	% letters a,b,c, same size as in text 
Figure 2. The periodic orbits of the fast variables.\ Points where interactions take place are indicated. If these occur in the orbit of a single fast variable (a), they will be difficult to miss, but in the case of two or more (b), the special points will be hit much less frequently, so that the interactions become slow. The orbit takes the shape of a (multidimensional) torus (c). \\ [5pt]

Note that energy is exactly conserved. Therefor we may assume that, if an initial state is dominated by the state \(|E=0\ket\), it will stay in that state.
 
Now consider the quantum model that we wish to mimic. Let that have a basis of \(N\) states, \(|\alf\ket,\,|\bet\ket,\ \cdots,\) with \(1\le \alf,\,\bet,\ \cdots <N\), to be called the \emph{slow} variables. Their interactions are introduced as \emph{classical} interactions with the fast variables, as follows:\bi{}
Two states \(|\alf\ket\) and \(\bet\ket\) are interchanged whenever the fast variables in the immediate vicinity of states \(\alf\) and \(\bet\)  simultaneously cross a certain pre-defined point on their (fast) orbits.\ei

 Here, the `vicinity' must be a well-defined notion for these states. In the case of non-relativistic particles, it means that we defined the states as the particle(s) in the coordinate representation \(|\vec x(t)\ket\). In the relativistic case we take the basis of states specified by the fields \(\phi_i(\vec x\,)\). This does imply that, in both cases, we regard the particles and/or fields to undergo exchange transitions that eventually will generate the desired Schr\"odinger equation or field equations.
 
 One can describe these classical interchange transitions in terms of a `quantum' perturbation Hamiltonian 
 \be H^\intt = \frac{\pi}2\sum_{\alf,\bet,s}\s_y^{[\alf,\bet]}\,\del_{\vv_\alf,\vv_\alf^{(s)}}\,\del_{\vv_\bet,\vv_\bet^{(s)}}\ , \eel{Hint1.eq}
 where \(\s_y^{[\alf,\bet]}\) is one of the three Pauli matrices \(\s_x,\,\s_y,\,\s_z\), acting on the 
 two-dimensional subspace spanned by the two states \(|\alf\ket\) and \(|\bet\ket\).
 
 Some special points on the orbits of the fast variables \(\vv_\alf\) and \(\vv_\bet\) will be indicated as
 \(\vv_\alf^{(s)}\) and \(\vv_\bet^{(s)}\). If the fast variables \(\vv_\alf\) and \(\vv_\bet\) reach their special positions \emph{simultaneously} then the corresponding classical states \(|\alf\ket\) and \(|\bet\ket\) are interchanged.
 
 In Eq.~\eqn{Hint1.eq}, we used a discretised notation, where the time unit is chosen such that it is the time needed to advance the fast variables by only one step in their (discretised) orbits. One may check that the factor \(\pi/2\) is crucial to guarantee that, if the special point is reached, the equation
 	\be \ex{-\fract{\pi i}2 \s_y}=-i\s_y=\twomat{\,0&-1}{1&0} \eel{exppauli.eq}
 describes a classical interchange, without generating superpositions. The minus sign is unavoidable but causes no harm. We chose the Pauli matrix \(\s_y\) because, when combined with the factor \(i\) in the Schr\"odinger equation, the wave function will be propagated as a \emph{real}-valued quantity.  One might desire to generate one of the other Pauli matrices also using classical physics. This can be done by adding a dummy binary variable, as described in Ref.~\cite{GtH2021} (the binary variable also propagates classically).
 
 The Hamiltonian describing the evolution of the slow variables is now derived by assuming that the fast variables never get enough energy to go to any of their excited energy states. Their lowest energy states are \(|0\ket^E\) obeying
 	\be {}^\ont\bra k|0\ket^E=\frac 1{\sqrt{N}} \ , \eel{zerostate.eq}
 so that the expectation value of a Kronecker delta is	
 	\be \bra 0|\del_{\vv_\alf,\vv_\alf^{(s)}}|0\ket=\frac 1 N\ , \eel{expvaluedelta.eq}
where \(N\) is the number of points on the fast orbit of this variable.

 Eq.~\eqn{exppauli.eq} could als be used if we had only one Kronecker delta in Eq.~\eqn{Hint1.eq}, but this would cause exactly one transition during one period of the fast variable, which makes the effective Hamiltonian too large to be useful. Choosing two Kronecker deltas causes one transition only to take place after much more time, making the insertion \eqn{Hint1.eq} of the desired order of magnitude to serve as a contribution in the \emph{effective} Hamiltonian of the slow variables.
  
 By adding a large number of similar transition events in the orbits of all fast variables, causing transitions for all pairs of (neighbouring) slow variables, we can now generate any desired contributions to the effective Hamiltonian elements \(H_{\alf\bet}\) causing transition among the slow variables. The result will be
 \be  H^\intt_{\alf\,\bet} = \frac{\pi}2 \s_y^{[\alf,\bet]}\, \frac{N^{[s]}}{N_{[\alf]}N_{[\bet]}}\ , \eel{Hint2.eq}
 where the numbers \(N_{[\alf]}\) and \(N_{[\bet]}\) are the total numbers of points on the orbits of the fast variables \(\alf\) and \(\bet\), and the numbers \(N^s\) indicate the numbers of the special transition points on the donut formed by the orbits of the pair \(\alf,\bet\).

 We encounter  the restriction that the matrix elements will come with  rational coefficients in front. The fundamental reason for the coefficients to be rational is that, eventually, all discretised classical models have finite Poincar\'e recursion times. In practice one may expect that this problem goes away when, for realistic classical systems, the Poincar\'e recursion times will rapidly go to infinity.
 
 We have now achieved the following. Let there be given any Hamiltonian with matrix elements \(H_{\alf\bet}\) in a finite-dimensional vector space, and given a suitably chosen basis in this vector space, preferably one where every state can be endowed with coordinates \(\vec x\). Then we have defined slow variables \(|\alf\ket\) describing the physical states, and we added fast variables whose excited states are beyond the reach of our experiments. We found classical interactions, prescribed as exchanges between the classical states, such that the effective Hamiltonian will approach the given one.
 
The system obeys the Schr\"odinger equation dictated by this Hamiltonian, and, by construction, all probabilities evolve as is mandated by the Copenhagen doctrine. The reader may ask how to obtain the diagonal elements of the Hamiltonian, and how to make it contain complex numbers. The answer is that these can also be generated by using an additional binary degree of freedom as mentioned above.

In principle, one could have used any orthonormal basis of states to be used in our construction, but in practice we would like to recover locality in some way. The demand of locality in the classical system implies that we should demand locality for the fast variables and the slow ones. This appears to be straightforward. For non-relativistic particles, one may use the basis of states defined by the position operators \(\vec x\). In the relativistic case, one needs the field operators \(\vv_i(\vec x)\) and their quantum eigen states to start off with.
 
The theory we arrive at appears to be closely related to Nelson's `stochastic quantum mechanics'\,\cite{Nelson1966}. We think our construction has a more solid mathematical foundation, explaining how the quantum entanglement arises naturally from the energy conservation law, associated to time translation invariance.
 
 \newsecl{Symmetries and superpositions}{symmsup.sec}
 
	The interpretation of the Schr\"odinger equation that we obtained is that it merely describes the evolution of the probability distribution for the slow variables, after averaging over the positions of the fast variables. The fast variables dictate the evolution, but they act too fast for us to observe this directly. The new thing in our procedure is that we have the choice to also describe the fast variables using quantum mechanics as a tool: fast and slow variables together go into a vector representation of what happens.\fn{Do keep in mind that the distinction between fast variables and slow variables is a feature of our simplified models, but possibly unnecessary in the real world. All variables are real, evolving classically according to the same or similar classical laws.}
	
	In statistical treatments of moving variables, with well-determined evolution laws, it should be completely clear that the probability distributions of the final state are the result of our choices for the probability distributions of the initial states.
	
	At first sight, the group of rotations can also be regarded as pure permutations, and, although the lattice structure of our local coordinate space \(\vec x\) will be severely affected, one might suspect that our present understanding of physics comes from smearing the lattice back into a continuum; this may be a reasonable approach towards understanding rotational symmetry. 
	
	However, more severe problems arise if we consider the notion of spin in a particle. We need to take spin into account when analysing Bell's theorem. In the treatment displayed in the previous section, the spin variable of a particle would be a discretised variable \(s\), with integer spacings ranging from \(-S\) to \(S\), where \(S\) is the total spin quantum number, being integer or half-odd-integer. These would be promoted to the status of classical variables, and then we can set up exactly the right Schr\"odinger equation for particles like the Dirac particle. What happens with its ontological interpretation if we rotate that?
	
	It is clear that, in this case, rotation transformations transform the `real states' \(|s\ket\) into superpositions. In doing so, the rotation group can serve as the prototype of many symmetry considerations in quantum mechanics. How do we analyse the Stern-Gerlach experiment?
	
	The superimposed states obey exactly the same Schr\"odinger equations as the basis elements do, and we had chosen the latter at will. So one possible answer could be: it does not matter which of the states we call real; there is no experiment to help us make the distinction. But this is debatable. The Stern-Gerlach experiment in its vertical orientation distinguishes particles with spin up from particles with spin down, these have different orbits. Remember that, in this chapter, we focus on going beyond the usual statistical interpretation of quantum mechanics, aiming at a description of pure, real states. The only accepted probabilities are 1 and 0.
	
 	The real physical states we work with form a basis of Hilbert space, and the equations we work with ensure that any state that starts off being real, occurring either with probability 1 or with probability 0, continues to be a real state  forever. This must also hold for any Stern-Gerlach set-up or any of the other paradoxical contraptions that have been proposed over the years. Real state in \(=\) real state out. This was called the `law of  ontology conservation'\,\cite{GtH2019}.
	
	At first sight it seems that a Stern-Gerlach experiment, after a rotation over an arbitrary angle, turns into a superposition of several real states. This is true in the mathematical sense. It is the easiest way to visualise what the rotation group stands for. However, if we physically rotate a Stern-Gerlach experiment, by undoing and re-arranging nuts and bolts, We do something else. The new experiment again goes into one of the realistic states; the nuts and bolts also go into new physical states, so this is not quite the same kind of rotation.
		
	Notice however, that if a  particle with spin leaves one Stern-Gerlach instrument and continues its way in an other, rotated, device, then, as we know from standard quantum mechanics, it emerges in a superposition, or more accurately, in a probabilistic distribution.  Where does this stochastic behavior come from? What happens if we do interference experiments with the various emerging beams of particles?
	
	Apparently, there are other variables that play a role. We can blame the fast variables for this. The fast variables for the rotated device do not coincide with the previous fast variables.   In specifying the state of the particle in the first device, we forgot to observe where exactly the fast variable was. We couldn't observe this, as it was moving too fast. The transformation (in this case that is the rotation), formally involved the excited energy modes of the fast variables.  In practice, we know that the energies of the quantum particles in both devices are too low to detect the excited modes, but in formulating the interactions, using the special points in Figure 2.b and c, the excited modes do play a role because the interaction points are localised.
	
	From these considerations, we claim that  whatever is left of the various paradoxes should be nothing to worry about.

\newsecl{On Bell's theorem}{Bell}	\def\source{{\mathrm{source}}}
	Yet, this conclusion is often criticised. To set the stage, let us recapitulate J.S.~Bell's theorem\,\cite{Bell1964}: a source is constructed that emits two entangled photons simultaneously. Such sources exist, so no further justification of its properties is asked for. If \(\pm z\) is the direction of the photons emitted, then the helicities are in the \(x\,y\) direction. Entanglement here means that the 2-photon state is\fn{Alternatively,  a source emitting two spin \(\halff\) particles could be used. The angles of the polarisation will then be twice the angles of the photon orientation discussed here, and there will be other modifications due to the fact that these particles are fermions.}
		\be|\psi\ket_\source = \fract 1{\sqrt 2}(|0\,0\ket+|1\,1\ket)\ , \eel{source.eq}
where the \(0\) stands for the \(x\) polarisation and \(1\) stands for the \(y\) polarisation.
Alice and Bob use polarisers to analyse the photons they see. Alice rotates her polariser to an angle \(a\) and Bob chooses an angle \(b\), and these choices are assumed to be totally independent. The two photons ``do not know" what the angles \(a\) and \(b\) are; they are  assumed to emerge with a polarisation angle \(\lab\). According to the usual view of what hidden variables are, the probability that both Alice and Bob are detecting their photon is written as \(P(a,b)\); the probability that a photon with orientation \(\lab\) is detected by Alice is assumed to be 
\(p_A(\lab,a)\), and the probability that Bob makes an observation is written as \(p_B(\lab,b)\). One then writes
	\be P(a,b)\qu\int\dd\lab\cdot\rho(\lab)\cdot p_A(a,\lab)\cdot p_B(b,\lab)\,. \eel{Probab.eq}
All probabilities \(p\) and \(P\) are assumed to be between \(0\) and \(1\).  \(\rho(\lab)\) is also positive and integrates to one. Bell would argue that this expression should apply to  theories such as ours, simply by merging the fast variables \(\vv_i(\vec x)\) with the parameter \(\lab\).  \\[10pt]
Figure 3.\\
Bell's definition of locality. 1 and 2 are two small \\
regions of space-time, space-like separated. \\
``Full specification of what happens in \\ 
region 3 makes events in 2 irrelevant \\ for predictions about 1 if \\
local causality holds". \\[-70 pt]
\hok\qqquad\qqquad\qqquad\qqquad\qquad	\widthfig{200pt}{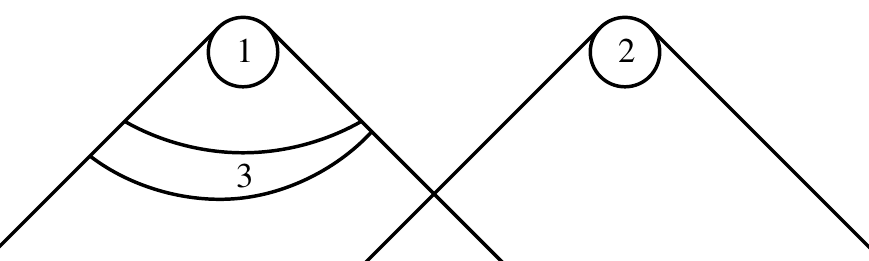}\\[10pt]

Figure 3 shows what assumptions go in Eq,~\eqn{Probab.eq}. It seems to be obvious that
observers in regions 1 and 2 may choose any setting \(a\) and \(b\) to identify their photon.

Writing \(\ol a=a\pm 90^\circ\),  photons obey:
	\be P(\ol a)+P(a)=1	\ . \ee
The correlation between Alice's and Bob's measurement is then written as 
	\be E(a,b)=P(a,b)+P(\ol a,\ol b)-P(a,\ol b)-P(\ol a,b)\ . \eel{correl.eq}
Standard quantum mechanics allows one to choose the entangled photon state \eqn{source.eq} as if it is oriented towards
either Alice or Bob, since it is rotation independent. The outcome is then 	\def\quant{{\mathrm{quant}}}
	\be E_\quant(a,b)=2\cos^2(a-b)-1=\cos\,2(a-b)\ . \eel{correlquant.eq}
In the fashionable hidden variable language, Eq.~\eqn{Probab.eq} is assumed to be valid, which implies that the photon 
must take care of giving Alice and Bob their measurement outcome whatever their choices \(a\) and \(b\) are, and these outcomes are found to obey the CHSH inequality\,\cite{CHSH1969}, derived directly from Eq.~\eqn{Probab.eq}. One then finds that Eq.~\eqn{correlquant.eq} conflicts with eq.~\eqn{Probab.eq}.  There is a mismatch of at least a factor \(\sqrt 2\), realised when 
\(|a-b|=22.5^\circ\) or \(67.5^\circ\).

Several `loopholes' were proposed, having to do with the limited accuracy of the experiments, but these will not help us, since we claim that our theory exactly reproduces quantum mechanics, and therefore Eq.~\eqn{correlquant.eq} should be reproduced by our theory. It violates CHSH. How can this happen?

Our short answer is that we have a classically evolving system that exactly reproduces the probability expressions predicted by the Schr\"odinger equation, including Eq.~\eqn{correlquant.eq}, in a given basis of Hilbert space. The model is local and allows for any initial state;  it does not require any kind of `conspiracy' or `retrocausality', or even non-locality. 

This should settle the matter, but it is true that the violation of the CHSH inequality is quite surprising.

The difficulty resides in assumption \eqn{Probab.eq}. Bell derives it directly from causality. If no signal can travel from the space-time point where Alice does her measurement to the point where Bob does his experiment, and \textit{vice versa}, then Eq.~\eqn{Probab.eq} just follows. Nevertheless, an assumption was made.

It amounts to the statement that the variables \(\lab,\ a,\) and \(b\), are mutually independent.
However, in Ref.\,\cite{GtHCA}, we computed the minimal non-vanishing correlations between the angles \(a\), \(b\) and \(\lab\) that could reproduce the quantum expression \eqn{correlquant.eq} exactly. We found\fn{This outcome is model dependent, and if we choose the model to be physically more plausible, the correlations become even stronger.}:
	\be P(a,b,\lab)=C|\sin\,2(a+b-2\lab)|\ , \eel{threecorrel.eq}
where \(C\) normalises the total probability to one (its value depends on the integration domain chosen). 
 This expression shows a non-vanishing 3-variable correlation, without any 2-variable correlations as soon as one averages over any of the three variables. An equation such as \eqn{threecorrel.eq} should replace \eqn{Probab.eq}.
 
One can read this to mean that the settings \(a\) and \(b\) have an effect on \(\lab\), but one can also say that  the choice of \(\lab\) made by the photon, affected the settings chosen by Alice and Bob.\fn{This then would be an example of the `butterfly effect'. It is not as crazy as it sounds. As soon as we include the fast variables in the discussion, the dynamics becomes invariant under time reversal, and the statement that a later photon is correlated with settings chosen earlier is then not strange at all.}  Perhaps the best way to interpret this strange feature is  that it be an aspect of information: the fact that the fast variables occupy all states in their orbits with equal probabilities is expressed by saying that they live in their energy ground states. The choice of the phases here is a man-made ambiguity that may propagate backwards in time. It is not an observable `spooky signal', since nothing propagates backwards in time in the classical formulation.

When we say that the photons (together with the fast variables) `affect' the settings chosen by Alice and Bob, it implies that Alice and Bob have no `free will'. Of course they haven't, their actions are completely controlled by the equations. We can't change setting \(a\) without changing what happens in region 3 of Figure 3.

It is important then to realise that our theory is \emph{not} a theory about statistical distributions. If we include the fast variables, everything that happens in region 3, occurs with probability \(1\), or, if it does not happen, it has probability \(0\). There is no in-between. Remember that we reproduce the Schr\"odinger equation \emph{in a given basis of Hilbert space}. The probabilities of the Schr\"odinger equation emerge exactly, but only if we start with the right basis elements.

We can add to this an important observation when the classical degrees of freedom are considered: \emph{even a minute change of the setting \(a\)  will require an initial state in Figure 3, region 3 that is orthogonal to what it was before that adjustment.} This is because the settings are classically described. The required rotation of the fast variable erases the information as to where its transition point was located (see Figure 2).

We note that this aspect of our scenario implies the absence of `free will' for Alice and Bob in choosing their settings. Alice and Bob are forced to obey classical laws, such that the rule \emph{ontology in \(=\) ontology out} is obeyed. The same can be said of Schr\"odinger's cat. Eventually, what we see when inspecting the cat is its classical behaviour. Only after adding the (in practice invisible) fast variables, we can perform a basis transformation to quantum states to say that the cat is superimposed. The statement belongs in the world of logic generated by the vector representation, but means nothing as long as we hold on to the classical description.

 \newsecl{Where are the fast variables and the slow variables in the Standard Model?}{SM.sec}

At first sight we may seem to be a long way from describing quantum field theories such as the Standard Model. In principle, one may expect something resembling a cellular automaton, where we may be able to project the various field variables as data on a cellular lattice. However, as described in section~\ref{symmsup.sec}, we have to deal with the question how continuous and discrete symmetry patterns, essential for the Standar Model to work, can be introduced. As is well-known, once we have all local and global symmetries in place, the entire Standard Model is almost fixed, with only a few dozen interaction parameters to be determined. We make a gentile attempt at finding some sign posts that could indicate to us where to start.

In a very important paper\,\cite{Jegerl}, F. Jegerlehner describes the Standard model as a minimalistic outcome of an algebraic structure whose basic interaction properties are essentially natural near its ultimate cut-off scale, the Planck length, except that the Higgs field self coupling happens to vanish, or almost vanish, at that scale. It seems as if the universe is metastable, or perhaps just at the edge of stability. When we scale towards the TeV scale, using the renormalization group equations, one discovers that the Higgs self-coupling slowly grows towards its present value, and this appears to explain the recently observed Higgs mass remarkably well. 

There are important new questions that may be raised in connection with the present work. One is where in the cellular automaton this copious algebra is generated; and of course we want to know how any kind of fast oscillating variables can arise. Previously, this author was just thinking of very heavy virtual particles such as the vector bosons that represent the remaining grand unification symmetry, but there is a problem with that:  as described in Section~\ref{realmodel.sec}, Figure 2, the dynamical fast variable must have the geometry 
of a multi-dimensional torus, whereas fields have a more trivial topological structure if they indeed form vector representations of the unifying algebra, see section~\ref{realmodel.sec}. 

This perhaps can be done better\,\cite{GtH2021b}. The general philosophy that might be useful here starts from a fundamental observation.  Fields that describe data at the Planck scale, can only propagate as fields at much more conventional scales (from milli-electronVolts to nearly a TeV), if there is a mechanism that prevents them from obtaining effective mass terms. To be precise, the dynamical field equations must allow them to be shifted by a constant with only minor effects on the energy of the state. At our scale of physics (to be referred to as the SM scale), fields can be shifted in any way, depending on space and time, such that energies also change within the energy domain of our SM scale. This means that the effective mass term must be at the SM scale. When we move towards the Planck scale, this mass term must rapidly approach to zero. Physically, the only mechanism that can do this is the Goldstone mechanism:

\emph{Only if a field effectively describes a symmetry transformation, and if, at the Planck scale, our world is invariant under this symmetry transformation, then we can understand how this field can propagate all the way to the SM scale.}

Since the Standard Model has a rich spectrum of possible fields (fermionic and bosonic), this would force us to suspect that each of these fields must represent a symmetry transformation under which the Planck-scale theory is either exactly invariant (when the mass term vanishes) or invariant in a very good approximation (when the mass is of SM scale or smaller). Indeed, this should also hold for the fermionic fields, and this points towards supersymmetry at the Planck scale.

In short, every field component in the SM represents a generator for an almost exact symmetry of the Planck scale model. If we would be dealing with only scalar fields in the Standard Model, this would give us all the symmetry transformations, including estimates on how well the system is invariant.

Unfortunately, the real situation will be a lot more complicated. We have fermionic fields that transform as spinors under rotation, and we have vector fields that themselves again obey local gauge symmetries. How do we deal with that? It would be a great assignment for a team of PhD students to design and elaborate a logically coherent mathematical scheme.

This scheme might eventually produce logical guidelines for setting up cellular automaton models in such a way that their behavior at SM scales indeed reproduce the SM. But this is not all. The resulting automaton will still be a \emph{quantum} automaton. What we now need is a set of variables that can play the role of fast variables. These are fields, but they cannot live on a flat field-space, they must form toruses as  in Section~\ref{realmodel.sec}. Now it would be tempting to consider the gauge groups. All group parameters of the local
gauge groups \(SU(2),\ SU(3)\), and \(U(1)\), form toroidal spaces or spheres. What's more, we know that the physical quantum states are invariant under these group transformations, so regardless their time-dependence, our world should be in the invariant state, just like the energy ground state. This could be an alley towards understanding how quantum behaviour could follow from a classical cellular automaton. 

\newsecl{General Relativity and black holes}{GRBH.sec}

Finally, there is General Relativity. This theory must be regarded as  just an other theme of the general concept of local gauge theories. It represents a non compact gauge group of curved coordinate transformations and it may well be that it can be handled similarly. It is important to remember that this theory is not renormalizable when presented in its usual form. We do observe that the addition of one further interaction term, the square of the conformal Weyl curvature term \(C_{\mu\nu\alf\bet}\), restores renormalisability at the cost of negative energy modes\,\cite{GtH2015}. Perhaps this mode can serve as a fast variable, but much more work will be needed to remedy various difficulties.

Theories for quantum mechanics that also aspire to include General Relativity, must address the fundamental black hole question. Black holes that are sufficiently large and heavy compared to the Planck scale of units, can be perfectly well described by classical, \emph{i.e.}  unquantised Einsteinian laws. However many researchers appear to arrive at the conclusion that there is something wrong with the black hole horizon, which might even involve the larger black holes. The origin of this suspicion is the emergence of `firewalls' forming a curtain of destruction against particles entering (or leaving) the horizon. The firewalls originate from the Hawking particles that are expected to emerge in the more distant future. 

The present author found that there exists a unique procedure to neutralise the firewalls, but it does not happen automatically. To see what may well happen, one should compute the effects that particles entering a black hole have on the Hawking particles leaving. It is not an act of destruction but a precisely calculable effect of repositioning those rays of out going material. The bottom line is that the \emph{positions} of the out going particles are effected by the \emph{momenta} of the in-going ones, and, because of quantum duality relating position to momentum, the same relation is found when going backwards in time: the momenta of the out-going particles are linked to the positions of the in-going ones.

These findings allow one to construct a unique expression for the \emph{black hole evolution matrix}, only requiring very basic knowledge of the mathematics of GR and QM.

However, we also hit upon a more sobering difficulty, The region behind the horizon has to be used to describe the \emph{time reverse} of the region normally visible, otherwise the evolution matrix (actually a quantum evolution matrix) fails completely to be unitary. For someone familiar with the Schwarzschild metric and its generalisations that have charge and angular momentum, there is no surprise here, but for the quantum physicist, this presents a problem. If we reverse the time direction, we also change the signs of all energies of the matter particles. Yet quantum field theories became successful precisely because they ensure the positivity of the energy of all particles. Can we allow ourselves a theory with such apparently conflicting properties?

The only answer that we could find is that we should act in a way similar to what P.A.M. Dirac did in order to overcome the negative energy problem im the Dirac equation. Now in a black hole, we have bosons and electrons alike, but we can achieve the same result by assuming that the entire band of energy eigenstates in a field theory should be bounded from below \emph{and from above!}  In that case, we can interpret the energy states beyond the horizon to be filled with particles completely, if the region at our side of the horizon is empty, and the other way around. The name \emph{anti-vacuum} was coined, describing the completely filled state. 

The region beyond the vacuum then represents a CPT inversion of the region at our side of the horizon. This picture appears to make perfectly sense, and we believe it to be likely that it resolves the energy inversion problem in black hole physics.

This solution of the energy inversion problem replaces the infinite energy spectrum of all harmonic oscillators generated by the fields outside the horizon, with a spectrum of evenly separated energy levels that have both a beginning and an end, the end being the highest possible energy level. We note that this is not only the energy spectrum of an atom with finite spin inside a homogeneous magnetic field (the Zeeman atom), but it also represents the energy levels of a periodic system with  finite time steps \(\del t\) in its evolution law, see the beginning of section \ref{realmodel.sec}.

Indeed, we find that black holes may be telling us something about the origin of quantum mechanics.

\newsecl{Conclusions}{conc.sec}
Our aim was to rescue the concept of ontology  as opposed to epistemology in quantum mechanics. This tells us that the atoms, molecules, electrons and other tiny entities are features of things that really exist. They evolve into different states or objects that also exist, according to universal physical laws. We find that this makes perfect sense if what we now perceive as quantum mechanics is understood as a vector representation of the states as they exist and evolve. Vector representations themselves allow superposition, and one finds that the superpositions of `ontological' states evolve through the same Schr\"odinger equations as the original states themselves. This in turn implies that one may ignore everything that is said about ontological existence as long as we use Born's dictum that the absolute squares of the superposition coefficients represent probabilities. The reason why we nevertheless attach much importance to our ontological interpretation is that it implies a severe restriction for the evolution laws; asking for the existence of an ontological representation forces us to redesign the set of elementary basis elements of Hilbert space, which might implicate new constraints on what kinds of Standard Model  we may suspect to describe our world.

An ontological interpretation is also of great help in resolving the numerous `paradoxes' that have been around confusing scientists as well as young students as to what `reality' really is about. Questions such as the physical process that seems to be associated to the `collapse of the wave function', the `measurement problem', as well as the difficulties raised in the EPR paper as well as Bell's theorem, questions surrounding the features of entanglement, and the Greenberger - Horne - Zeilinger  (GHZ)  paradox, all become much less counter intuitive and mysterious than what they look like in their original quantum settings.

	The explanation of these features is that the real thing that is happening is the classically evolving collection of microscopic objects, of which the fastest periodically moving things automatically enter into a completely featureless, even distribution over all of their possible states.

	Remarkably, the reason why the states of the fastest moving objects stay in an even distribution is better understood in the quantum formalism than when using the original classical picture: the highest energy excitations are difficult or almost impossible to excite, simply because the energy needed for that is usually unavailable to us: in our accelerators we can only reach a dozen or so TeVs, and in cosmic rays the highest detectable energies are still well below the Planck scale. Therefore, the excited modes are only virtually present, and may well be ignored in practice. And, since all superposition coefficients for the ground state are equal, the distribution is featureless, in practice -- according to Born.
	
	Thus, what we really find is that the lowest energy states of the slow variables become entangled due to their interactions with the fastest variables. Quantum mechanics ensues; it is mathematically inevitable.
	
	Our work is far from finished. Fresh young minds should probe the remaining mysteries; in particular, the Standard Model is built from fundamental symmetry principles. There are more symmetries than one might have expected from `just any' classical system: there are many continuous symmetries, and also non-compact symmetries such as Lorentz invariance and general coordinate transformation invariance, and there are exact local gauge invariances as dictated by the gauge fields in the Standard Model.
	
	Finally, a natural place must be found where we can put and understand the black hole solutions of Einstein's equations. They too must obey the laws of quantum mechanics, before we can embrace these remarkable systems in our overall picture of nature. Data obtained from the observations of cosmologists must also be incorporated. What we are searching for is nothing less than a grand picture of the evolution laws shaping our physical world.

 \end{document}